\documentclass{article}

\pdfoutput=1
\usepackage{PRIMEarxiv}
\usepackage[utf8]{inputenc} 
\usepackage[T1]{fontenc}    
\usepackage[hidelinks,urlcolor=blue]{hyperref}
\usepackage{url}            
\usepackage{booktabs}       
\usepackage{amsfonts}       
\usepackage{nicefrac}       
\usepackage{microtype}      
\usepackage{lipsum}
\usepackage{fancyhdr}       
\usepackage{graphicx}       
\graphicspath{{media/}}     
\usepackage{textcomp}
\begin{document}

\title{Public-private funding models in open source software development: A case study on \textit{scikit-learn}}

\author{
  Cailean Osborne \\
  Oxford Internet Institute \\
  University of Oxford \\
  Oxford, UK\\
  \texttt{cailean.osborne@oii.ox.ac.uk}
}

\maketitle

\begin{abstract}
Governments are increasingly funding open source software (OSS) development to support software security, digital sovereignty, and national competitiveness in science and innovation, amongst others. However, little is known about how OSS developers evaluate the relative benefits and drawbacks of governmental funding for OSS. This study explores this question through a case study on \textit{scikit-learn}, a Python library for machine learning, funded by public research grants, commercial sponsorship, micro-donations, and a €32 million grant announced in France's artificial intelligence strategy. Through 25 interviews with \textit{scikit-learn}'s maintainers and funders, this study makes two key contributions. First, it contributes empirical findings about the benefits and drawbacks of public and private funding in an impactful OSS project, and the governance protocols employed by the maintainers to balance the diverse interests of their community and funders. Second, it offers practical lessons on funding for OSS developers, governments, and companies based on the experience of \textit{scikit-learn}. The paper concludes with key recommendations for practitioners and future research directions. 
\end{abstract}

\keywords{Open source software, OSS funding models, machine learning, scikit-learn, OSS sustainability}

\section{Introduction}
Open source software (OSS) has moved up the priority list of governments in light of concerns about digital sovereignty \cite{osborne_european_2023}, software security \cite{herpig_fostering_2023}, and national competitiveness in science and innovation \cite{nagle_government_2019}, amongst others. The discovery of the \textit{log4j} vulnerability in November 2021 was a turning point that animated discussion about OSS funding. From the White House's  software security meeting in January 2022 \cite{white_house_readout_2022} to the establishment of the Sovereign Tech Fund (STF) in Germany in October 2022 \cite{stf_sovereign_2022}, there has been a growing recognition amongst policymakers of the public value of funding the development and maintenance of OSS, increasingly recognised as digital infrastructure underpinning our digital society and economy \cite{keller_european_2022,eghbal_roads_2016}. Increasing governmental interest and involvement in funding OSS has been generally well-received by OSS developers. However, little is known about how they view the relative benefits and drawbacks of governmental funding compared to other funding sources that have funded OSS developers to date.

This study explores this question through a case study on \textit{scikit-learn}, a Python library for machine learning (ML), which is described as ``the Swiss army knife of ML'' due to its widespread use in research and industry \cite{inria_2019_2020}. Its sustainability as a community-led project in the industry-dominated field of artificial intelligence (AI) \cite{ahmed_growing_2023} is partly attributed to its mixed funding model, including public research grants, commercial sponsorship, community donations, and the institutional backing from the French Institute for Research in Computer Science and Automation (Inria). In November 2021, in its national AI strategy, la Stratégie IA, the French government announced a €32 million grant to support the expansion of \textit{scikit-learn} \cite{sgpi_france_2021}, marking a significant expansion of public funding for \textit{scikit-learn}.

Through 25 interviews with \textit{scikit-learn}'s maintainers and funders, this study investigates the historical role of public and private funding in supporting the community-led OSS project. This study makes two key contributions to research and practice. First, it contributes empirical findings about the  benefits and drawbacks of public and private funding in an OSS project. Furthermore, it sheds light on how the maintainers employed governance protocols to balance the diverse interests of their funders and to safeguard their community ethos. Second, it offers practical lessons on funding in community-led OSS projects for OSS developers, companies, and governments. 


The paper has the following structure. First, it discusses prior work on the role of funding in supporting the sustainability of OSS projects. Second, it presents the study design and the case of \textit{scikit-learn}. Then, it presents the results and discusses their implications for research and practice. The paper concludes with a discussion of the key recommendations to practitioners and suggestions for future research directions on OSS funding models.

\section{Related Work}
\subsection{The Role of Funding in OSS Sustainability}

The adoption of OSS is ubiquitous, estimated to be used in 96\% of code bases \cite{synopsys_open_2023} and constituting up to 90\% of software stacks \cite{open_source_security_foundation_oss_2022}. As a result, OSS is considered a critical component of the digital infrastructure that underpins modern society and the digital economy \cite{eghbal_roads_2016,panezi_open_2020,scott_avoiding_2023}. Given its importance, the health and sustainability of OSS projects is a concern for a diverse range of stakeholders. However, the health and sustainability of OSS projects is a complex and multi-faceted issue \cite{goggins_open_2021,lumbard_empirical_2024,xiao_how_2023}. A range of factors are relevant, including but not limited to open governance and strong leadership \cite{omahony_boundary_2008,trinkenreich_hidden_2020, li_leadership_2012, huang_identifying_2022}; maintainer best practices, such as timely responses \cite{salkever_open_2023}, mentoring contributors \cite{tan_is_2023, zhou_who_2015}, and recommending tasks \cite{xiao_recommending_2022}; a steady inflow of new contributors \cite{avelino_abandonment_2019, barcomb_uncovering_2020, zhou_who_2015}; and diversity within the community \cite{bosu_diversity_2019}. Amongst these factors, the financial health of a project plays a crucial, but hitherto understudied, role in the sustainability of OSS projects, determining the extent to which maintenance is funded or reliant on volunteers \cite{eghbal_working_2020}.

However, money has long been a contentious subject amongst OSS developers. On the one hand, many developers take pride in the origins of the OSS movement as a social movement \cite{broca_communs_2021}, describing it as a ``programmers' paradise'' for ``geeks'', ``hackers'', and ``hobbyists'', who do not contribute to OSS for money \cite{raymond_cathedral_2001,zhang_who_2021}. Developers cite their political ideals \cite{kelty_two_2008}, altruism \cite{markus_what_2000}, or their passion for spotting and quashing bugs in software \cite{loebbecke_open_2003} as intrinsic incentives for contributing to OSS. Furthermore, many developers complain that ``money ruins everything'' \cite{eghbal_where_2017}. On the other hand, a plethora of OSS projects are unfunded and maintained by overstretched volunteers, who often struggle to keep up with the workload required to keep projects running and communities thriving \cite{eghbal_working_2020,geiger_labor_2021,salkever_open_2023}.

The consequences of systemic issues of free-riding and under-investment in OSS are highlighted when developers spot major vulnerabilities that pose significant threats to digital systems in governments, critical infrastructure, and companies' products and services. In particular, the discovery of vulnerabilities, such as the \textit{heartbleed} bug in April 2014 or \textit{log4j} vulnerability in December 2021, which was described as the ``worst security problem in a generation'' \cite{vaughan-nichols_hard_2021}, shocked industry and governments alike. It highlighted the need for cross-sector stakeholders, including the public and private sectors, to fund the developer communities who build and maintain our open source digital infrastructure \cite{keller_european_2022}. As Benjamin Bikinbine has argued, there is an urgent need ``not just [for] investment in institutions, organisations, technologies, or innovations, but long-term and sustainable investment in the true source of their value: people'' \cite{birkinbine_incorporating_2020}.

\subsection{Private Funding for OSS Development}
The private sector has been the largest funder of OSS to date, funding OSS developers and projects in both direct and indirect ways. The impact of commercial investments in OSS are gauged to be significant. According to one estimate, companies in the European Union (EU) invested around €1 billion in OSS in 2018, which generated between €65-95 billion for the EU's GDP that year \cite{blind_impact_2021}. Direct funding methods include sponsorship of OSS foundations and consortia, which operate on a fee-paying membership model \cite{omahony_emergence_2007}. Foundations and consortia are known to play a critical role as ``boundary organisations'', which through their open governance protocols facilitate collaboration between diverse volunteers and companies \cite{omahony_boundary_2008}. However, according to a survey of 369 OSS developers, only 12\% of respondents reported that organisational funding for OSS via a foundation, a consortium, or an independent legal entity was either extremely or very useful to their development activity, while 5\% reported it was ineffective \cite{tidelift_2020_2020}.

Recently, companies have begun to establish FOSS Contributor Funds, which allow employees to nominate and vote on OSS projects that should receive funding \cite{obrien_sustaining_2019}, and represent a more democratic way for companies to fund the OSS projects that they use \cite{obrien_foss_2019}. A shortcoming is that these funds typically award up to \$10,000 over a one-year time frame. Alyssa Wright from Bloomberg's FOSS Contributor Fund \cite{wright_bloomberg_2023} has acknowledged this shortcoming, arguing that ``true sustainability comes from a multi-sided commitment'' and ``while financial support is important, real engagement through contribution and dialogue is as crucial to the ecosystem’s survival as writing a check.''

Other ways companies support OSS projects by sponsoring OSS developers or by allowing their employees to contribute to OSS projects at work \cite{butler_company_2021,xia_lessons_2023}. Sponsoring developers is a recognised strategy that companies employ to influence projects \cite{dahlander_how_2008,dahlander_man_2006} and to improve their reputation as OSS patrons, which in turn is considered to aid the recruitment of developers \cite{bonaccorsi_comparing_2006,chesbrough_measuring_2023,pitt_penguins_2006}. In the mid-2000s, it was estimated that 40\% of OSS contributors were paid \cite{lakhani_why_2003}. We should expect this number to be much higher today, with companies like Google, Microsoft, and Amazon accounting for a rapidly growing number of OSS developers and contributions to OSS projects \cite{hale_google_2022}. In addition, companies invest in projects through donations and by joining project steering committees \cite{butler_investigation_2018}. 

While OSS projects benefit from commercial participation, maintainers seek to manage the influence or the dominance of individual companies \cite{zhang_corporate_2022}. A popular strategy is to establish or transfer projects to foundations \cite{omahony_emergence_2007,wagstrom_vertical_2009}, which through open governance protocols limit the dominance of single companies \cite{di_giacomo_key_2020} and facilitate collaboration between diverse contributors \cite{omahony_boundary_2008,germonprez_open_2013}. Crucially, these organisations enable OSS developers to ``negotiate their relationship with companies and, when necessary, to defend their commons-based resources from unwanted influence'' \cite{birkinbine_incorporating_2020}. 

\subsection{Micro-Donations for OSS Development}
Individuals or organisations support OSS developers and project by making micro-donations to individual developers and OSS projects. For instance, the GitHub Sponsors scheme, launched in 2019, allows users to make one-off or recurring payments to individuals or organisations with sponsored profiles \cite{github_sponsoring_2023}. According to prior work, common incentives for donations via GitHub Sponsors include expressing appreciation, recognising a developer’s work, and encouraging future contributions \cite{zhang_who_2021}. By comparison, common incentives for maintainers to accept donations are to receive recognition for their work or to provide a side-income; while those who do not participate in GitHub Sponsors state that they do not need to be sponsored or that they do not contribute to OSS for money \cite{zhang_who_2021}. Furthermore, tther work shows that the amount that individuals donate is associated with the length of their use of an OSS \cite{krishnamurthy_monetary_2009}. The impact of micro-donations is still contested. On the one hand, they have been shown to shorten response times to issues \cite{nakasai_analysis_2017,nakasai_are_2019} and to increase maintenance-related activities \cite{medappa_sponsorship_2023}. On the other hand, overall the impact appears trivial \cite{overney_how_2020} and donations tend to be ad-hoc, small, or legally complicated for individuals to receive \cite{eghbal_handy_2022}.

\subsection{Public Funding for OSS Development}
The public sector has indirectly or directly funded OSS since the origins of the World Wide Web \cite{cern_birth_2023} through research grants for (scientific) OSS \cite{during_trouble_2006, howison_scientific_2011,strasser_ten_2022} as well as bespoke OSS funding bodies, such as the Open Technology Fund in the United States of America (USA), the Next Generation Initiative at the European Commission, or the STF in Germany \cite{keller_european_2022}. Furthermore, OSS has recently moved up the priority list of governments in light of concerns about the security of OSS \cite{vaughan-nichols_log4shell_2021,herpig_fostering_2023}; its role as critical digital infrastructure  \cite{eghbal_roads_2016,panezi_open_2020,scott_avoiding_2023}; and its importance for digital sovereignty \cite{osborne_european_2023,burwell_digital_2022} and national competitiveness in science and innovation \cite{nagle_government_2019}. In Europe, the concern for digital sovereignty is particularly salient. For example, the European Commission's OSS strategy argues that OSS can give ``Europe a chance to create and maintain its own, independent digital approach and stay in control of its processes''  \cite{european_commission_european_2020}. 

Public sector support for OSS is also a recognised strategy to stimulate competition in domestic software markets, which might otherwise be monopolised by industry giants \cite{jokonya_investigating_2015}. For example, it has been estimated that the\textit{ Circulaire 5608}, a French law requiring government agencies to favour OSS when procuring software, led to an increase of nearly 600,000 OSS contributions from French developers per year as well as yearly increases in the national competitiveness of the French IT market \cite{nagle_government_2019}. Beyond governmental adoption of OSS, government investments in OSS are often targeted at stimulating national competitiveness. For example, in June 2023, President Emmanuel Macron of France announced a €40 million fund to create a digital commons to support the development of open source large language models to support domestic startups and to challenge the dominance of Big Technology companies in the AI industry \cite{chatterjee_france_2023}.

The discovery of the \textit{log4j} vulnerability in November 2021 marked a turning point for governmental interest and involvement in funding OSS development. In light of the magnitude of the security concerns \cite{vaughan-nichols_log4shell_2021}, governments rapidly mobilised to address the threats posed by OSS vulnerabilities to their critical infrastructure and digital economies. For example, in January 2022, the White House convened cross-sector stakeholders, including government agencies, Big Technology companies, and OSS foundations, to discuss the urgent need to improve the security of OSS, recognising its indispensable role as digital infrastructure whose maintenance largely depends on volunteers \cite{white_house_readout_2022}. Across the pond, in June 2022, the newly formed European Working Team on the Digital Commons, comprising representatives from 19 member states of the EU and the European Commission, published its inaugural report, highlighting the importance of funding OSS for bolstering the digital sovereignty of Europe \cite{european_working_team_on_digital_commons_towards_2022}. In October 2022, Germany established the STF, allocating a budget of €11.5 million for its first year alone to fund OSS maintenance in the interest of digital sovereignty, software security, innovation, and digital democracy \cite{stf_sovereign_2022}. These developments underscore the growing recognition amongst policymakers of the public value of funding OSS development and maintenance \cite{keller_european_2022}.

While there is a growing recognition of the need for sustainable funding for OSS by governments, little is known about how OSS developers evaluate the relative merits and drawbacks of governmental involvement in funding OSS development compared to other types of funding that have primarily supported OSS development so far, or the impact thereof on OSS development and maintenance. This study addresses this gap by examining the case of scikit-learn and how its maintainers view the benefits and drawbacks of its public and private funders. By focusing on the perspectives of OSS developers, this study contributes practical insights on this timely question in the debate about OSS sustainability.

\section{Study Design}

\subsection{Research Aims}
This study investigates the emerging role that governments are playing in funding OSS development and maintenance, and how OSS developers view the relative merits and drawbacks of governmental funding through a case study on the \textit{scikit-learn} project. In software engineering research, a case study ``is an empirical enquiry that draws on multiple sources of evidence to investigate one instance (or a small number of instances) of a contemporary software engineering phenomenon within its real-life context'' \cite{runeson_case_2012}. 
Single case studies are suitable when one seeks to provide in-depth insights, generate new hypotheses, or develop theories about a critical, extreme, unusual, or revelatory phenomenon \cite{yin_case_2018}. The \textit{scikit-learn} project was selected as an exploratory case that could provide in-depth insights into the role of the French government in supporting this OSS project and how its maintainers view the relative merits and drawbacks of the government’s funding compared to the other funding sources that have supported the project to date. 

To this end, the study focuses on the following research questions (RQ):
\begin{itemize}
    \item \textbf{RQ1}: What are the interests of the public and private funders of \textit{scikit-learn}, and how have they aligned or conflicted with the interests of the \textit{scikit-learn} maintainers?  
    \item \textbf{RQ2}: How do the maintainers of \textit{scikit-learn} view the relative benefits and drawbacks of public and private funding for their OSS project?
    \item \textbf{RQ3}: What practical lessons can be drawn from \textit{scikit-learn}’s funding model for OSS practitioners?
\end{itemize}

\subsection{Case Presentation}
\textit{scikit-learn} is a Python library that implements ML algorithms for classification, regression, and clustering, as well as tools for data pre-processing, model evaluations, and data visualisation. Initially it was started as \textit{scikits.learn} by David Courapeau during a Google Summer of Code project in 2007. After a dormant period, it was relaunched as \textit{scikit-learn} by Fabian Pedregosa, Gaël Varoquaux, Alexandre Gramfort, and Vincent Michel with the backing of Inria, with the first public release in February 2010. It has since become one of the most popular OSS libraries for ML, described as ``the Swiss army knife of ML'' due to its widespread use \cite{inria_2019_2020}. For example, in November 2023, it had 2,752 contributors and 61,085 GitHub stars \cite{oss_insight_oss_2023}. The success and sustainability of this community-led OSS project is noteworthy, considering the extent of industry dominance in ML/AI research \cite{ahmed_growing_2023} and OSS ecosystems \cite{langenkamp_how_2022,white_model_2024}.

\textit{scikit-learn}'s sustainability is partly credited to its mixed funding model \cite{scikit-learn_scikit-learn_2023, varoquaux_foundation_2018}. First, Inria has funded \textit{scikit-learn} since 2010, including by paying salaries, administering research grants, and providing office space, amongst others. Second, community members have made micro-donations via NUMFOCUS and student projects have been sponsored by the Google Summer of Code programme. Third, the \textit{scikit-learn} consortium, established under the Inria Foundation in 2018, offers three levels of annual memberships for companies: silver (€30,000), gold (€50,000), and platinum (€100,000), each providing varying degrees of involvement in the Technical and Advisory Committees. The Technical Committee develops the strategic technical roadmap for the project, which involves collecting input from the community, while the Advisory Committee provides guidance on strategic matters and the consortium's membership, amongst others. Since its founding, it has included Dataiku, Microsoft, Nvidia, Intel, AXA, Boston Consulting Group, BNP Paribas Cardif, Hugging Face, Nvidia, Fujitsu, and Chanel. Companies, including Quansight Labs, Nvidia, and HuggingFace, have also sponsored maintainers. Fourth, in November 2021, the French AI strategy, la Stratégie IA, announced a €32 million grant to support \textit{scikit-learn} \cite{sgpi_france_2021}. This significant public investment, alongside other funding sources, makes \textit{scikit-learn} a suitable case for the research aims of this case study.

\subsection{Data \& Analysis}
\subsubsection{Data Collection}
To answer the research questions, this study employs a mixed-methods approach, comprising document analysis and 25 semi-structured interviews with \textit{scikit-learn} maintainers, commercial sponsors, and government officials between November 2021 and April 2023. Most interviews took place during on-site visits to the \textit{scikit-learn} team's office in Paris. The sampling approach sought to ensure complete coverage of the various stakeholder groups in the project---that is, the OSS developers (i.e. the \textit{scikit-learn} maintainers), commercial sponsors (i.e. the consortium members), and public funders (i.e. Inria and the French government)---as well as diversity within each group. Interviews were conducted with maintainers with various tenures in the project, including some who had been involved for over 10 years and recent university graduates. One maintainer worked for a company that was a consortium member but spoke on his own behalf. From the consortium, interviews were conducted with three representatives from three companies with different sponsorship tiers, organisational sizes, and different sectors. In addition, interviews were conducted with two government officials who were responsible for the coordination of the AI strategy. The interviews were semi-structured, combining general questions about \textit{scikit-learn}’s funding model and tailored questions for each interviewee \cite{kvale_interviews_1996}. To inform tailored questions for each maintainer, their contribution histories on GitHub’s Contributor Insights Page were reviewed; and for the funders, public information about their involvement in funding \textit{scikit-learn} was reviewed. The author conducted and recorded every interview to aid the analysis and enhance the validity of the findings \cite{yin_case_2018}.

The interview data was supplemented and triangulated through detailed field notes from on-site visits and secondary documents to strengthen the validity of the findings \cite{yin_case_2018}. The field notes stemmed from two on-site visits to the \textit{scikit-learn} office in Paris-Saclay in March 2022 and 2023; two on-site visits to the offices of sponsors in September and December 2022; and attendance at the Global Partnership on AI Summit in November 2021, which was hosted by the French government and attended by a co-founder of \textit{scikit-learn}. The secondary documents included both publicly available and non-disclosed documents, which interviewees shared concerning the French government’s AI Strategy and its funding package for \textit{scikit-learn}. Additional information about \textit{scikit-learn} and its funding model was collected from internet searches, included the \textit{scikit-learn} website and company blogs, or provided by interviewees.

\subsubsection{Qualitative Data Analysis}
Throughout the 17 months of data collection, an iterative, integrated approach was employed to identify common themes in the interview data, field notes, and secondary documents \cite{braun_using_2006}. Given the exploratory aims of this study, the integrated approach combined the application of grounded theory analysis \cite{charmaz_constructing_2006} and a deductive, organising framework informed by prior literature about the different interests and involvement models of individuals, companies, and governments in the development of OSS  \cite{birkinbine_incorporating_2020,bonaccorsi_comparing_2006,jokonya_investigating_2015,omahony_boundary_2008}. Thematic codes were continuously developed as data was collected, and in turn translated into themes until the point of saturation \cite{cruzes_recommended_2011}. The resulting themes were member-checked to ensure their accuracy and resonance with research participants \cite{lincoln_naturalistic_1985}. Throughout the research process, a social identity map was maintained by the author to exercise reflexivity regarding the potential influence of personal biases stemming from the author's views and relationships built with the \textit{scikit-learn} team during the 17 months of data collection \cite{jacobson_social_2019}.

It should be noted that each stakeholder group was not assumed to be homogenous; instead, it was accepted that there could be divergent or even conflicting interests within each stakeholder group. For instance, the \textit{scikit-learn} consortium comprises a range of companies, including startups and multinational corporations from different sectors. Prior work indicates that companies primarily pursue strategic interests by participating in OSS \cite{bonaccorsi_entry_2006}, and it was not assumed that the consortium members shared the same strategies in funding \textit{scikit-learn}. Similarly, public funders, including Inria and government officials who were responsible for France’s AI strategy, were grouped together solely by their belonging to the French public sector, despite obvious differences between these stakeholders. Even the maintainers were a heterogeneous group, combining employees of Inria and external companies, each with their unique paths to and histories in the \textit{scikit-learn} project. These theoretical groupings were used solely for analytical purposes and were iteratively refined during the 17 months of data collection to ensure their coherence and suitability for the analysis. 

\section{Results}
The results are organised into three parts, sequentially responding to \textbf{RQ1} on how funders' interests align with or differ from those of maintainers and \textbf{RQ2} on the relative benefits and drawbacks of public and private funding from the perspective of the maintainers. Finally, practical lessons and recommendations for practitioners, as per \textbf{RQ3}, are discussed in the subsequent Discussion section.

\subsection{Stewards of the Community: The Roles \& Interests of the \textit{scikit-learn} Maintainers}
\textit{scikit-learn's} maintainers are a diverse group employed by Inria, public universities, and software companies. While some core team members work on \textit{scikit-learn} full-time, others balance maintainer roles with their day jobs. Several maintainers also hold academic positions. The maintainers explained they perform a range of technical and social tasks, including reviewing code, mentoring contributors, and community building. François Goupil, the community manager, explained, ``Of course, software development itself is the major part of the workload---going through issues, pull requests, and responding to community feedback---but lots of other work goes into the project. Community animation is a huge part.'' In addition, company-sponsored developers make focused contributions, such as the development of new features, which are agreed upon by the maintainers and the respective companies. For example, Nvidia sponsors a full-time maintainer, who is building a plug-in to enable alternative computational backends for \textit{scikit-learn} to enable compatibility with Nvidia's GPU AI hardware accelerators.

Several maintainers expressed a strong sense of stewardship for the project's diverse community, comprising contributors from across the world (see Table~\ref{tab:sklearn-country}) and companies (see Table~\ref{tab:sklearn-company}). Gaël Varoquaux, a co-founder, explained that the community was at the heart of the project's DNA and underlined the importance of taking a community-driven approach to major decisions about the future of the project. Julien Jerphanion, a maintainer, highlighted the importance of community contributions, ``The major part of the labour [in the project] is based on benevolence of people working in their free time and not asking to get paid.'' Meanwhile Adrin Jalali, a maintainer, commented on the impact of micro-donations from community members, explaining that while ``the major stuff is not funded through NUMFOCUS'', it has been useful in funding marketing, events, and an internship for underrepresented groups.

In addition, several maintainers expressed their scientific interest in ML and the project's scientific culture as important incentives for contributing to \textit{scikit-learn}. Many maintainers had academic backgrounds, worked in research labs, or taught at universities. For example, Jerphanion said that he enjoyed working at the intersection of statistical theory and software engineering, which was lacking in his previous industry job. In addition, an anonymous maintainer explained that while he may not become a millionaire, he had a comfortable life and enjoyed the job's challenge. Furthermore, for many, being a \textit{scikit-learn} maintainer was a source of pride, given \textit{scikit-learn}'s global reputation and impact.

Beyond maintaining \textit{scikit-learn}, several maintainers discussed the importance of contributing to their dependencies, such as \textit{numpy} and \textit{scipy}, and the wider scientific Python ecosystem. 
Goupil explained that they also collaborate with these projects by applying for grants together. In addition, some maintainers had initiated satellite projects, such as \textit{joblib} and \textit{skrub}. However, their ability to support the ecosystem is constrained by limited capacity. Jerphanion explained that while projects like \textit{scikit-learn}, \textit{numpy}, and \textit{scipy} are very important in ML/AI, ``There are many which are not visible to the public, for example OpenBLAS which is fundamental and only maintained by two people, who have very specific expertise. This is tremendous software work, but it is invisible to the public. Projects like this are cornerstones for many projects. They are the foundations; they are the infrastructure.''

\begin{table*}[h]
    \centering
    \caption{Most Active Countries in the \textit{scikit-learn} Community \cite{oss_insight_oss_2023}}
    \small
    \begin{tabular}{p{0.5cm}p{4cm}p{4cm}p{4cm}}
        ~ & \textbf{Stargazer} & \textbf{Issue Creator }& \textbf{Pull Request Creator} \\ \toprule
        1 & China (23.7\%) & USA (33.7\%) & USA (34.0\%) \\ 
        2 & USA (21.8\%) & Germany (7.5\%) & India (8.9\%) \\ 
        3 & India (7.9\%) & UK (7.0\%) & Germany (8.5\%) \\ 
        4 & Germany (4.4\%) & India (6.8\%) & France (6.9\%) \\ 
        5 & Brazil (3.8\%) & France (5.5\%) & UK (6.4\%) \\ 
        6 & UK (3.7\%) & China (5.4\%) & Canada (4.5\%) \\ 
        7 & Canada (3.3\%) & Canada (4.5\%) & China (2.6\%) \\ 
        8 & France (3.2\%) & Netherlands (2.5\%) & Japan (2.5\%) \\ 
        9 & Japan (2.5\%) & Switzerland (2.2\%) & Switzerland (2.3\%) \\ 
        10 & Russia (1.3\%) & Australia (2.0\%) & Netherlands (2.2\%) \\ \bottomrule
    \end{tabular}
    \label{tab:sklearn-country}
\end{table*}

\begin{table*}[h]
    \centering
    \caption{Most Active Companies in the \textit{scikit-learn} Community \cite{oss_insight_oss_2023}}
    \small
    \begin{tabular}{p{0.5cm}p{4cm}p{4cm}p{4cm}}
        ~ & \textbf{Stargazer} & \textbf{Issue Creator }& \textbf{Pull Request Creator} \\ \toprule
        1 & Google (1.5\%) & Google (2.5\%) & Google (3.7\%) \\ 
        2 & Microsoft (1.5\%) & Microsoft (2.0\%) & Microsoft (2.5\%) \\ 
        3 & Tencent (0.9\%) & Amazon (0.7\%) & Inria (0.9\%) \\ 
        4 & Alibaba (0.8\%) & DeepMind (0.7\%) & Meta (0.8\%) \\ 
        5 & Amazon (0.8\%) & IBM (0.6\%) & Amazon (0.8\%) \\ 
        6 & ByteDance (0.8\%) & Johns Hopkins (0.6\%) & Dataiku (0.7\%) \\ 
        7 & Baidu (0.6\%) & Tencent (0.5\%) & UC Berkeley (0.6\%) \\ 
        8 & Meta (0.5\%) & ETH Zurich (0.5\%) & Shopify (0.6\%) \\ 
        9 & AWS (0.4\%) & Uni. of Washington (0.4\%) & MIT (0.6\%) \\ 
        10 & Tsinghua University (0.4\%) & NeuroData (0.4\%) & AWS (0.6\%) \\ \bottomrule
    \end{tabular}
    \label{tab:sklearn-company}
\end{table*}
\clearpage
\subsection{Public Funding: From Research Grants at Inria to la Stratégie IA} Public funding has supported the development of \textit{scikit-learn} since its beginning in 2010. Over 13 years since its first public release, Inria has provided an estimated €1.5 million through staff salaries, research grants, office space, computing resources, and event sponsorship. This led Olivier Grisel, a maintainer, to explain that, ``[Public funding] is not new with the AI strategy.'' However, the €32 million grant in France's AI strategy, la Stratégie IA, supporting the expansion of \textit{scikit-learn} and the development of OSS tools for data science, marked a significant expansion of public funding. The deputy coordinator of the AI strategy explained that the grant seeks to facilitate AI adoption, enhance French competitiveness in AI R\&D, and support the digital sovereignty of France and Europe.

The maintainers were delighted to receive the recognition from the government. Jalali highlighted the stability it provided for the project, enabling long-term planning around hiring and the technical roadmap. Jerphanion saw it as ``a good signal that the French state is getting involved in funding projects like this.'' Goupil especially commended the requirement to seek matching funds from public and private sources across Europe, viewing it as an important safeguard for the project's independence: ``I think it would be dangerous for us to be exclusively funded by the private sector or to be exclusively funded by the French government, because we have many good contributors who are not French.'' 

At the same time, challenges have emerged in aligning broad policy goals with the day-to-day realities and expertise of the maintainers. Initial requests from the government officials to expand \textit{scikit-learn} into deep learning to provide a ``sovereign'' alternative to PyTorch and TensorFlow were resisted, with the maintainers insisting on focusing on the project's specialism in ML and the need to allocate sufficient funding to maintenance. The coordinators of the AI strategy deferred to their expertise, with one official acknowledging, ``We understand that \textit{scikit-learn} is specialised in ML, and it should not compete against TensorFlow and PyTorch.''

However, other demands have proven trickier. Requirements to simultaneously develop OSS tools for data science whilst not competing with domestic companies have left maintainers in a bind, as the resulting tools will inherently provide free alternatives to the proprietary offerings of companies. Maintainers have also expressed frustration with the slow pace of disbursement, with funds still not arriving 17 months after the announcement by the time of the last interview. While \textit{scikit-learn}'s diverse funding model provided a buffer, they worried that such sluggishness could threaten the sustainability of more financially precarious OSS projects.

One of the government officials explained that the decision to fund \textit{scikit-learn}, as well as how to fund it, was informed by an ongoing multi-stakeholder process. ``It was a dialogue between several actors---with the AI strategy team, the `Confiance IA' project,\footnote{``Confiance IA'' is a government-sponsored programme that aims to develop trustworthy AI technologies in France.} Inria, with \textit{scikit-learn}, and so on.'' He confessed they had faced hurdles within the government: ``Financing open source means financing a product that cannot be valued on the market and \textit{a priori} it is not clear what its success will be. So, if [we want] the government to finance it in a substantial way---here we are talking about €32 million---the government wants to know where the money is going and to achieve a measurable return on investment.'' Ultimately, it took months of coordination to get the grant approved by the finance ministry.

Two consortium members raised concerns that such a substantial grant could politicise \textit{scikit-learn}, potentially influencing it to prioritise French policy goals over the interests of the community. One representative underlined the need to balance the French government’s concern about digital sovereignty with the norms of the global community, while another argued that imposing French policy priorities on the project would not be received well by the community. These concerns were acknowledged by the maintainers, who insisted that the community ethos of the project remained at the heart of their discussions with the government officials. 

\subsection{Private Sponsorship: The Role of the \textit{scikit-learn} Consortium}

Companies in the consortium have pursued a range of social, economic, and technological goals. Goupil explained that companies typically join the consortium to support the maintenance of \textit{scikit-learn} which they use in their products or services, as well as to build goodwill in the community, which can be useful for hiring ML researchers and developers. He commented, ``A lot of people come from this open source background and if you see that a company is promoting such an open source project, you might say, `Okay, maybe, I would be happy working with them. They could be okay with me contributing to OSS at work.''' He suggested this incentive may be more important for companies that do not already have a strong reputation in ML or AI, making them more credible and attractive to job-seekers.

For their part, consortium members cited a mix of pragmatic and principled reasons for funding \textit{scikit-learn}. On a basic level, it allows them to support the maintenance of a tool they use and rely on. Mike McCarty from Nvidia shared that his concern about unmaintained dependencies ``keeps [him] up at night,'' citing \textit{numpy} as a critical library that had been under-maintained for some time. Leo Dreyfus-Schmidt from Dataiku expressed a similar sentiment when he stated that, ``We are very much aware that we rely on open tools like \textit{scikit-learn}'' and highlighted the importance of ``giv[ing] back to the community.'' An anonymous consortium member was even more direct: ``Companies pay millions in commercial licenses, so why not support the OSS projects we use?''

Of course, consortium membership is not simply a charitable contribution. In exchange for their sponsorship, companies gain a voice in \textit{scikit-learn}'s development through bodies like the Technical Committee and Advisory Board, which are venues for surfacing industry use cases and needs. Dreyfus-Schmidt characterised the interactions in these committees as a ``win-win'' for the maintainers and the companies, allowing \textit{scikit-learn} to benefit from ``the perspective of people in industry, which...is a very different type of feedback from the one they get from the open source community,'' whilst giving commercial members direct input into the project.

In practice, this dynamic has yielded synergies and tensions. Largely, companies' interests have been compatible with those of the maintainers. For example, Nvidia has sponsored a maintainer to build a plug-in enabling alternative computational backends for \textit{scikit-learn}. McCarty explained, ``Nvidia is very interested in \textit{scikit-learn}'s effort to accelerate with GPUs. I would love to see \textit{scikit-learn} accelerated natively with a nice user experience, like \textit{PyTorch}, which you can install and accelerate without changing code.'' Guillaume Lemaitre, a maintainer, described Nvidia's simultaneous sponsorship and consortium membership as ``a win-win situation'' for the maintainers. However, in other cases, sponsors' requests to shape the project's direction or to burnish their reputations have run afoul of project norms. Two maintainers explained that a sponsor had once requested modifications to the technical documentation, which amounted in their view to free advertising of their products, which was unacceptable to the maintainers. While disputes of this kind have been rare, a dispute had led to the termination of a consortium membership in the past.

To balance the diverse interests of their community and the member companies in the consortium, the \textit{scikit-learn} maintainers employ governance protocols that limit funders' influence and prioritise community control. The Technical Committee, for instance, is barred from overriding decisions that achieve rough consensus amongst the maintainers. Jalali explained, ``We don’t want the technical committee to have power over the core contributor team as long as [the maintainers] have a consensus. If there’s two thirds majority in cast vote, for whatever vote we do, then the technical committee doesn’t step in…because the power is given to the [maintainers] and that’s by design.'' 

These governance measures are important to maintain the trust of the community and to mitigate fears that the project has been bought out. Goupil explained that as a community-led OSS project, they communicate to consortium members that their feedback is valued but they cannot guarantee that they will implement it. For example, the maintainers organised a workshop on MLOps because this was a topic that the consortium members had expressed an interest in. ``MLOps will probably end up in the technical roadmap in some way...but it’s a discussion and in the end the team will decide if it’s worth it or not,'' he explained. Through these governance protocols, the maintainers ensure that the project remains community-driven while still considering and incorporating feedback from the consortium. As Goupil put it, ``We are a community with different interests,'' and preserving that pluralism is key to the project's long-term sustainability and independence.

Finally, even though relationships with companies can be troublesome at times, the maintainers were generally grateful for their support and for not simply free-riding on \textit{scikit-learn}, which many companies do. For example, during an on-site visit, Goupil showed me a dashboard, comprising charts of contributions from companies. Upon observing that the most active organisations are Chinese and American companies (see Table~\ref{tab:sklearn-company}), I asked if these insights had influenced his opinion about whether they contribute sufficiently to the project. He responded frankly: while such companies contribute useful issues and pull requests, at the end of the day their contributions create more work for the maintainers. He explained that it would be more helpful if these companies sponsored a maintainer or joined the consortium in order to provide more capacity to the maintainers to process the extra workload created by these contributions.

\section{Discussion}
\subsection{Implications for Research}

\subsubsection{Relative Benefits \& Drawbacks of OSS Funding Sources}
This study contributes novel findings on how OSS developers view the relative benefits and drawbacks of public and private funding. In the case of \textit{scikit-learn}, private funding has been associated with an understanding of the importance of maintenance and it has introduced the maintainers to industry perspectives and use cases, but it has also required careful management of sponsors' interests. By contrast, the sizeable government grant has enabled long-term planning around hiring and the technical roadmap, but it has posed challenges in aligning policy goals with OSS realities, balancing national interests with community norms, and navigating the slow pace of disbursement. Overall, this case study shows that a diversified funding model can be an effective strategy for balancing the relative benefits and drawbacks of public and private funding. As Goupil explained, ``Overall, it’s good to have a panaché.'' 

The study also contributes novel insights on the diverse interests of OSS funders, from the economic and technological goals of companies \cite{bonaccorsi_comparing_2006} to the policy objectives of governments like digital sovereignty \cite{osborne_european_2023}, national competitive in science and innovation \cite{nagle_government_2019}, and economic growth \cite{jokonya_investigating_2015}. While public and private funders pursue different goals, the case study illustrates that they are not necessarily in conflict, sharing the goal of sustaining \textit{scikit-learn}. For example, while the maintainers and the companies do not necessarily share the government’s policy goals, such as digital sovereignty, its long-term investment in \textit{scikit-learn} aligns with the maintainers’ and companies’ interest in its sustainability for research and innovation purposes. Similarly, while the maintainers and the government officials do not necessarily share the companies’ business interests in \textit{scikit-learn}, the project benefits from companies’ feedback and funding for feature development and maintenance. 


\subsubsection{Designing Future Research on OSS Funding}
There is a dearth of research on OSS funding, relative to other topics on the economics of OSS \cite{cao_oss_2022}. With new funding models emerging, from FOSS Contributor Funds to governmental funds like Germany's STF, it is timely to empirically investigate OSS funding from multiple angles, including funders, incentives, funding models, and impact(s). Future research should expand this study by investigating the design and outcomes of different governmental OSS funding models, such as Germany's STF or the European Commission's Next Generation Internet initiative. 
By examining these aspects, researchers can contribute to the development of best practices and guidelines for the design and implementation of governmental funding programmes that effectively support the sustainability of OSS projects.  

\subsection{Implications for Practice}
\subsubsection{Recommendations for OSS Developers}
The case of \textit{scikit-learn} shows that a diversified funding model can be an effective strategy for safeguarding a project's independence whilst benefiting from both public and private support. Furthermore, public and private funding come with their unique benefits and drawbacks. On the one hand, private funders understand the importance of maintenance and share industry use cases that lead to the development of new features. However, their demands can conflict with project goals or norms. On the other hand, public funding can provide sizeable grants, which support long-term planning but may prioritise innovation over maintenance or policy goals over the specialism or needs of the project. Maintainers must be prepared to advocate for and defend their project's needs and community values, which are aided by project governance protocols and, if possible, regular dialogue during the design of grants. A diverse funding portfolio can provide both stability and leverage in these discussions.

\subsubsection{Recommendations for Companies}
For companies, the findings serve as a timely reminder that sponsoring OSS maintainers or projects can make a substantial difference to maintainers, who often struggle with limited capacity and towering workloads \cite{eghbal_working_2020}. As shown in Table~\ref{tab:sklearn-company}, several major technology companies, such as Google, Meta, and AWS, have benefited significantly from \textit{scikit-learn} but have not yet directly funded the project or its maintainers. While these companies may contribute useful issues and pull requests, their engagement can also create additional work for already overstretched maintainers. Sponsoring a maintainer or joining a project's funding consortium, even at a relatively low level, could be a high-impact way for well-resourced companies to support the OSS projects that they use and depend on. By stepping up to provide more direct financial support, companies can help ensure the sustainability and health of critical OSS infrastructure.

\subsubsection{Recommendations for Governments}
For governments, the findings extend prior recommendations for the funding of scientific OSS \cite{strasser_ten_2022} by highlighting the importance of funding the maintenance of existing OSS in addition to new innovation. To ensure that funding is aligned with project needs, governments should engage in multi-stakeholder dialogue with OSS developers. This consultation is crucial for designing grants that balance policy goals with the expertise of maintainers. In the case of \textit{scikit-learn}, initial government requests around expanding into deep learning were misaligned with the project's scope and capabilities. Ongoing conversations were necessary to bridge this gap and ensure that funding supported both overarching policy objectives and essential project functions. Governments should proactively seek input from the OSS community to avoid misalignments and to craft funding programs that effectively support the needs of critical OSS projects.

\subsection{Threats to Validity}
The validity of the study is evaluated by following guidance for case studies and qualitative research methods in the field of software engineering research \cite{runeson_guidelines_2008,yin_case_2018, easterbrook_selecting_2008}.

\subsubsection{Credibility}
Credibility refers to the believability of qualitative research findings \cite{easterbrook_selecting_2008}. Multiple data sources---interviews, fieldwork, and secondary documents—were combined to triangulate findings and minimise bias \cite{yin_case_2018}. The prolonged engagement with the project's maintainers over 17 months provided in-depth insights, minimising the biases of snapshots at one moment in time. Findings were member-checked with participants and reviewed by the community manager to enhance accuracy \cite{lincoln_naturalistic_1985, yin_case_2018}. However, data collection was constrained by access challenges, especially with high-ranking stakeholders in the government and sponsor companies, which is a common problem in ``elite interviewing'' \cite{mikecz_interviewing_2012}. For example, despite the outreach support of the community manager, only three representatives of commercial sponsors could be recruited for interviews. Despite these limitations, the triangulation of multiple data sources, prolonged engagement, and member checking helped to ensure the credibility and trustworthiness of the findings.

\subsubsection{Robustness}
Robustness concerns the strength, reliability, and soundness of the study's design, methods, and findings. A number of steps were taken to enhance the robustness. First, before the analysis, the theoretically-informed stakeholder groups were tested with interviewees to ensure their relevance, whilst accounting for nuances within and between stakeholder groups. This ensured that theoretical constructs were not imposed on the study that did not resonate with the research participants. Second, since the research data was collected over a prolonged time span of 17 months, the author took an iterative approach to the data analysis, combining an inductive, grounded theory analysis with a deductive approach based on the study’s theoretical framework throughout the period of data collection \cite{cruzes_recommended_2011}. Third, the author maintained a social identity map throughout the research process to minimise the influence of personal biases on the study \cite{jacobson_social_2019}

\subsubsection{Transferability}
Transferability concerns the generalisability of the findings. The single case study on \textit{scikit-learn} threatens the generalisability of the findings. Certainly, it is easier to generalise from multiple case studies, which by design test direct replication \cite{eisenhardt_building_1989} and mitigate the risk of case uniqueness \cite{herriott_multisite_1983}. For example, there are a number of unique characteristics of the \textit{scikit-learn} project, such as the longstanding institutional support of Inria, the design of the consortium, and the specific support from the French government. However, these threats to the validity of this study can be tempered by considering that,  contrary to statistical generalisation, case studies aim for analytical generalisation \cite{yin_case_2018}. For example, there are aspects of the \textit{scikit-learn} project that will resonate across OSS projects, including the maintainers’ experience of weaving the diverse interests of stakeholders into the project whilst seeking to safeguard the project’s community ethos, as well as the effective design and implementation of funding governance protocols. Furthermore, this study provides insights into the merits and challenges involved in the implementation of a public-private funding model in a community-led OSS project that may be useful for other projects considering similar approaches.

\subsubsection{Dependability}
Dependability refers to the consistency of the research process. To maximise the dependability of the research process, a systematic study protocol was employed for the collection, analysis, and storage of research data. Where consent was given, interviews were recorded and transcribed to aid analysis. When interview responses were ambiguous, follow-up discussions with the interviewees were initiated to obtain clarifications. In addition to interview data, detailed field notes from on-site visits and supplementary secondary documents facilitated data triangulation. Finally, the case study underwent thorough review, benefiting from the feedback from domain experts as well as peer review.

\section{Conclusion}
This study on the funding model of \textit{scikit-learn} makes two key contributions to research and practice. First, it contributes novel findings on emerging OSS funding models, illustrating how the maintainers of an impactful community-led OSS project have integrated and balanced public and private funding to sustain the development and maintenance of the project. Second, it offers practical lessons for the funding of community-led OSS projects, and it makes recommendations to various practitioners. In particular, it provides insights into the benefits and drawbacks of public and private funding for OSS, and how these different funding sources can be balanced to support project goals. Importantly, it highlights effectiveness of diversified funding models as a method to sustain community-led OSS projects in industry-dominated R\&D fields like AI. Overall, the findings contribute to the literature on OSS funding and provide actionable recommendations for governments and companies to support the sustainability of OSS.

\section*{Acknowledgements}
This study was funded by the UK's Economic and Social Research Council [Grant number: ES/P000649/1]. The author thanks the \textit{scikit-learn} maintainers and funders for participating in this study. 

\bibliographystyle{unsrt}  
\bibliography{manuscript}
\end{document}